\documentclass[aps, reprint, superscriptaddress, prl]{revtex4-2}
\usepackage{graphicx}
\usepackage{amssymb}
\usepackage{amsmath}
\usepackage{bm}
\usepackage{hyperref}
\usepackage{subcaption}
\usepackage{dcolumn}

\begin{document}
\title{Helicity-Selective Electron-Phonon Coupling Between $PT$-Partner States in $2H$-bilayer MoS$_2$}
\author{Fred Sun}
\email{fredsunqiushi@gmail.com}
\affiliation{BASIS Independent Silicon Valley, 1290 Parkmoor Ave, San Jose, CA 95126}
\author{Jia-An Yan}
\email{jiaanyan@gmail.com}
\affiliation{Department of Physics, Astronomy, and Geosciences, Towson University,
8000 York Road, Towson, MD 21252, USA}
\date{\today}

\begin{abstract}
Rotational symmetry ($C_n$) constrains chiral phonons, but how other symmetries affect their interactions remains largely unexplored. We perform a theoretical and computational study of chiral phonon coupling to parity-time
($PT$)-degenerate electron doublets in $2H$-bilayer MoS$_2$. Although $PT$ suppresses phonon singlet chirality, chiral phonon doublets can helically interact with $PT$-partner electronic states via a mechanism similar to pseudospin.
Selection rules are then analyzed with the interplay of $PT$ and $C_3$. Our computational results of $2H$-bilayer MoS$_2$ suggest chiral phonons may be used to resolve intravalley electronic states in $PT$ systems.
\end{abstract}

\maketitle

\textit{Introduction} -- Phonon chirality has become an active area of investigation once phonons were shown to be circularly polarizable and carry angular momentum \cite{Zhang_2014,Zhang_2015,Zhu_2018,Juraschek_2025, Park_2020}.
These chiral phonons can be characterized under $C_n$ rotational symmetries of the crystal lattice, generating a conserved quantum number known as pseudo-angular momentum (PAM) \cite{Streib_2021, Zhang_2022}. Being symmetry derived, this PAM enforces selection rules
for phonon interactions \cite{Pan_2024}. One such interaction is the electron-phonon coupling (EPC), a well-known mechanism underlying many physical phenomena \cite{Giustino_2017, zhang2026advancesphononsbandtopology}; in this, chiral phonons exhibit unique features by tuning the rotational degree of freedom in a system.
Chiral phonons have now been shown to generate electronic orbital angular momentum \cite{Yao_2026,Sato_2026}, spin magnetization \cite{Yao_2025}, and phonon magnetic moments \cite{Tang_2024, Xue_2025, Chaudhary_2024, Mustafa_2025}.

The relevance of other symmetries, in particular parity (P) and time-reversal (T) symmetry, or in brief, $PT$, in promoting chirality and chiral interactions, though, has only gained little attention. In many cases, $PT$ symmetry has been described as producing phonon achirality, since for phonon singlets, it forces angular
momentum to be zero \cite{Hang_2018}. More recently, $PT$ has been shown to produce finite-momentum chiral phonons via $PT$-partnered degenerate doublets at an arbitrary crystal momentum \cite{Zhang_2026}. This, however, is not a necessary condition for doubly degenerate
chiral phonons at $\Gamma$ since time-reversal ($T$) can relate $C_n$-induced degenerate doublets. Moreover, the effects of $PT$ symmetry on chiral phonon interactions, in particular with electrons, whose intravalley scattering is mediated by $\Gamma$-phonons, remains unclear. Indeed, $PT$ introduces
a new subtlety in interaction space since electron states become doubly degenerate, which makes it difficult to apply existing chiral selection rules.

\begin{figure}
    \centering
    \includegraphics[width=\linewidth]{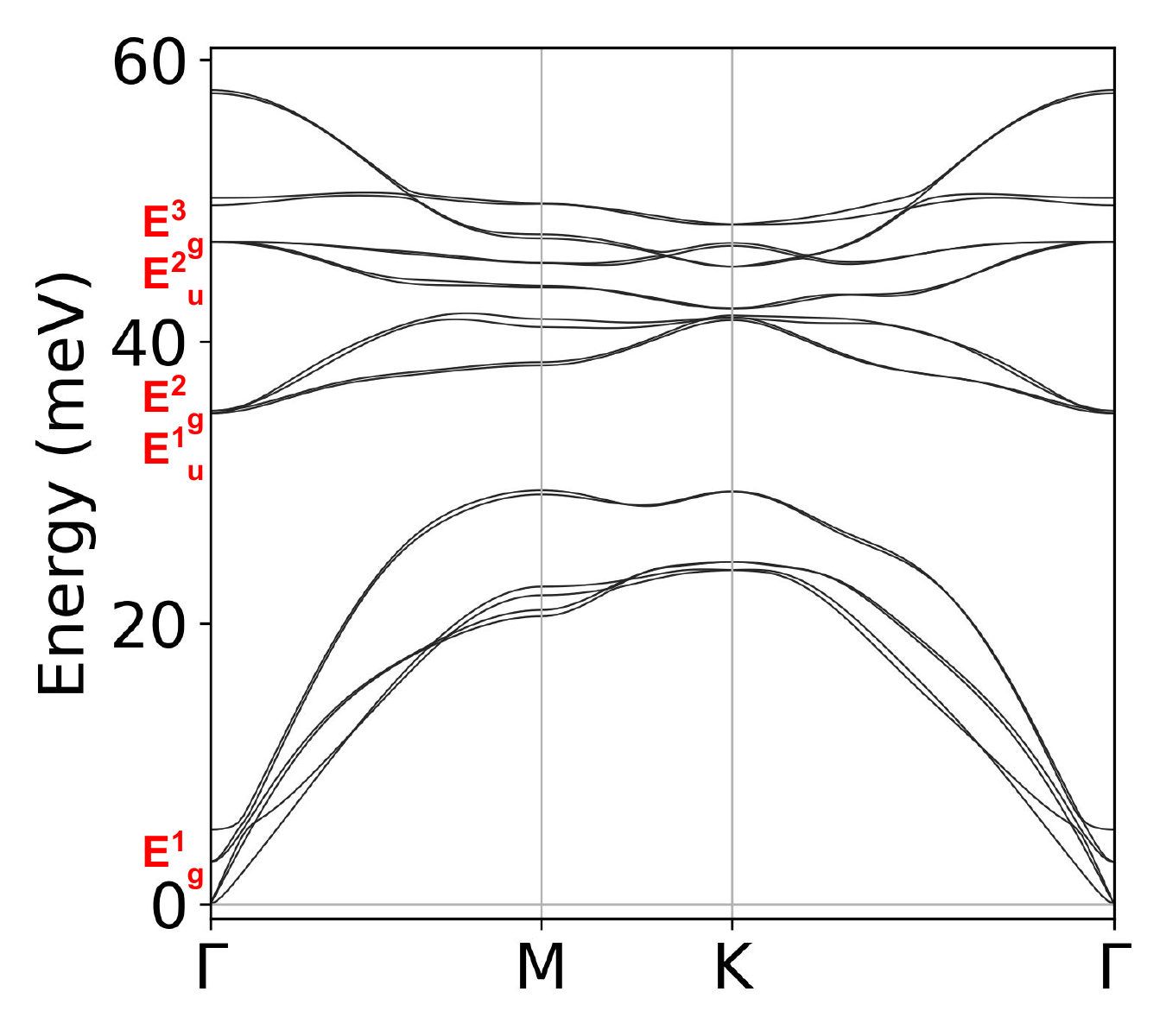}
    \caption{Phonon dispersion for $2H$-bilayer MoS$_2$, with the 2D symmetry irreps labeled in red. We see that $2H$ hosts several doubly degenerate irreps at $\Gamma$, for which chiral combinations may be formed
    and the helicity-selective coupling to the $PT$-induced electron pseudospin may be observed.}
    \label{fig:2H-phonon-dispersion}
\end{figure}

We derive that these $\Gamma$-phonons can interact chirally with these electrons, the chiral structure of which is mediated by $PT$. Indeed, $PT$ symmetry is shown to induce a band pseudospin on the electrons for which the chiral phonons can select.
We study this phonon chirality in the context of the $C_3$-symmetric $2H$-bilayer MoS$_2$, but our derivation is applicable to general $PT$ $C_n$ systems. The importance of $C_n$ symmetry here is to apply it to the resultant electron-phonon structure
derived from $PT$ which allows one to ascertain the chiral selection rules. In this, they apply to $C_n$ invariant electronic $k$ points. These selection rules reduce the coupling matrices to at most rank one, wherein opposite phonon helicities can select for $PT$ partners. Our numerical analysis confirms these chiral signatures in $2H$-bilayer MoS$_2$,
whose phonon band dispersion is plotted in Fig. \ref{fig:2H-phonon-dispersion}, notably containing several doubly degenerate phonon irreducible representations (irreps) at $\Gamma$. This allows one to construct the chiral phonon combinations which exhibit dark electron transition coupling channels of zero coupling strength. Between
opposite helicity combinations, these dark transition states are $PT$-partners (and by orthogonality, the bright transition states are also $PT$-partners).
Importantly, this transition is intravalley due to the phonons being at $\Gamma$, meaning the constructed pseudospin is contained inside that valley, as opposed to the well-known $K/K'$ valley pair pseudospin related by $T$. We thus propose chiral phonons as a promising
method to resolve intravalley electronic states in many $PT$ systems.

\begin{figure}
    \centering
    \includegraphics[width=\linewidth]{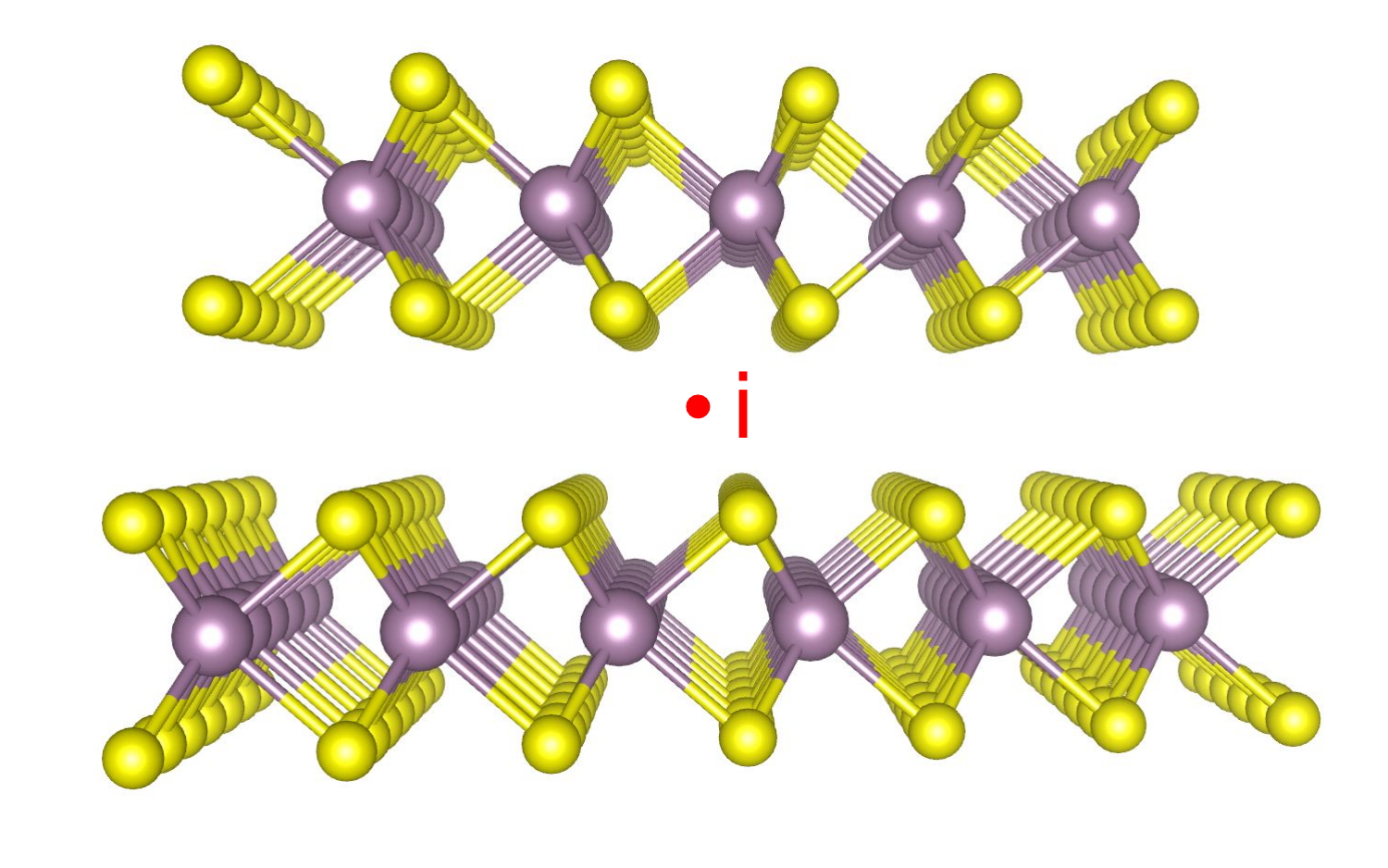}
    \caption{Atomic structure of 2H-bilayer MoS$_2$. Notably, the lattice contains an inversion center, an example of which we have denoted with a red point. }
    \label{fig:2H-inversion}
\end{figure}

\textit{Methods} -- Density functional theory (DFT) and density functional perturbation theory (DFPT) calculations were performed using the Quantum ESPRESSO (QE) code \cite{Giannozzi_2009} with PBE exchange-correlation functional. Fully relativistic optimized norm-conserving Vanderbilt pseudopotentials were adopted \cite{Hamann_2013}, allowing us to perform calculations including spin-orbit coupling (SOC). A $12\times12\times1$ k-point grid and
plane wave cutoff energy of $70$ Ry were used. A large enough vacuum space ($>$15 \AA) was also included along the perpendicular direction in the supercell to minimize the fictitious image interactions. EPC elements were calculated with the EPW code \cite{Giustino_2007, Lee_2023}. All atomistic models were created using VESTA \cite{Vesta, Vesta_3}.
The $C_3$ axis is chosen to intersect the inversion center, so that $PT$ commutes with $C_3$.

\begin{figure*}
    \centering
    \includegraphics[width=\linewidth]{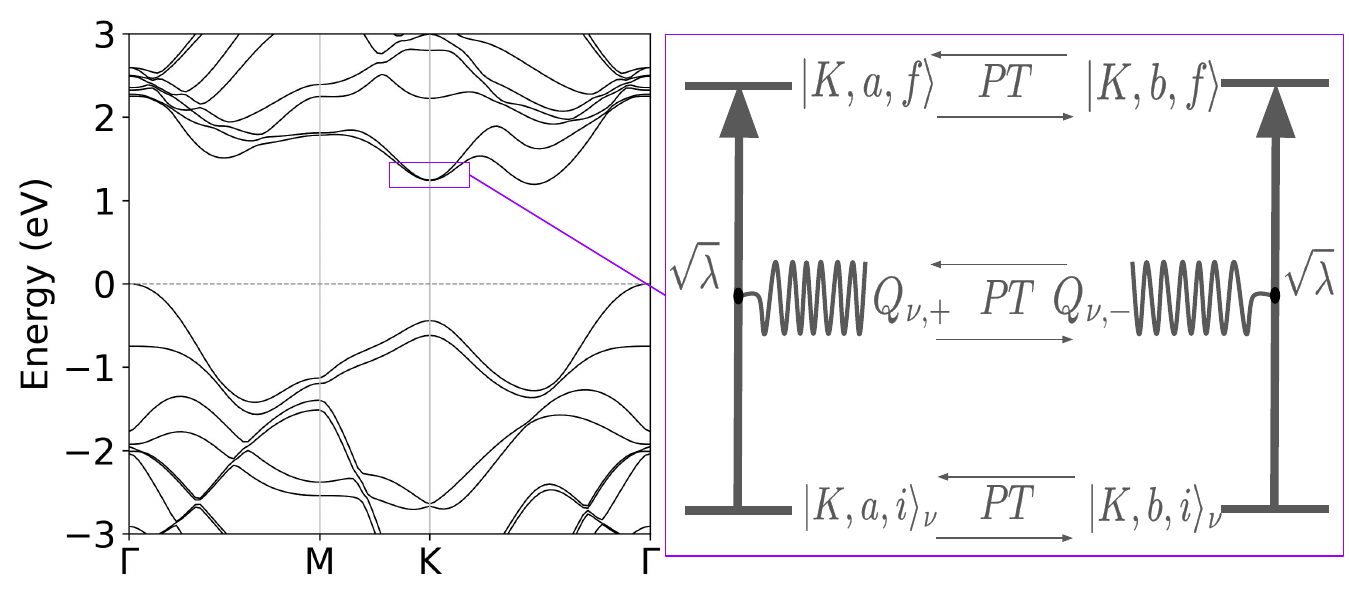}
    \caption{(Left) Band structure of $2H$-bilayer MoS$_2$, wherein we have highlighted and given a schematic for the chiral $PT$-based electron-phonon interaction (right). The valence band maximum has been shifted to zero. The selected electron states that couple to the $Q_{\nu, +}$ and $Q_{\nu, -}$ phonons are labeled by $a, b$ respectively and are related by $PT$ within an initial/final doublet. Moreover,
    chiral selectivity shows that $Q_{\nu, -}, Q_{\nu, +}$ have zero coupling with $a, b$ states, respectively. In $2H$ bilayer MoS$_2$, the initial conduction band states are dependent on phonon mode, since the initial electron subspace has real $C_3$ eigenvalues and therefore
    are not constrained by rotational symmetry. The square root of the transition strength eigenvalue $\lambda$ is analogous to EPC elements for a two-dimensional interaction space.}
    \label{fig:2H-transition}
\end{figure*}

\textit{$PT$-symmetry and chirality} -- Fig. \ref{fig:2H-inversion} shows the structure of $2H$-bilayer MoS$_2$. Indeed, it contains an inversion center and thus has parity ($P$) symmetry. Moreover, since the system is not magnetic, we have time-reversal ($T$) symmetry, for a combined
$PT$ symmetry. On phonon singlets, $PT$ on the phonon itself gives the condition on PAM that $\ell = -\ell$, forcing it to be zero. However, another realization of phonon angular momentum is within the $C_3$ induced doubly degenerate
phonon subspaces. Choosing a chiral phonon basis $Q_{\nu, \pm} = \frac{Q_{\nu, x} \mp iQ_{\nu, y}}{\sqrt{2}}$ \cite{Chaudhary_2024}, the two phonons are related by $PT$ and have opposite helicities $\ell = \pm 1$. Even more, the electronic space of the $2H$-bilayer is also doubly degenerate, since
$PT$ maps electronic crystal momentum back to itself and exhibits a Kramers-like orthogonal degeneracy relation \cite{Loss_2009}. Define $O$ as the representation of $PT$ on a given state space (phonon, initial electronic, final electronic, etc.),
where we will for convenience omit subscripts to label $O$ when unambiguous. Then, choosing an electronic state $|a\rangle$ inside this space, we can define its $PT$ partner $|b\rangle = O|a\rangle$ which is guaranteed to be orthogonal and degenerate because
$(PT)^2 = -1$ on electrons as fermions.
Hence, $a,b$ label a kind of electron pseudospin induced by $PT$ symmetry, though it should be noted that currently the exact state $|a\rangle$ is arbitrary and will only become constrained once we find the transition matrices. Still, the relation between
$|a\rangle, |b\rangle$ suggests a kind of pseudospin physics, so we will define the electron pseudospinor $\Psi = (c_a, c_b)^T$. The relevant phonon is the $\Gamma$-point one, to conserve crystal momentum. Then, for a given phonon mode, the relevant terms in the EPC Hamiltonian can be compactly written as
\begin{equation}
    H'_{\Gamma K} =  \boldsymbol{\Psi}_f^\dagger(G_-Q_+ + G_+Q_-)\boldsymbol{\Psi}_i+\mathrm{h.c.},
\end{equation}
where the coupling constants become $2\times 2$ matrices to reflect the size of the interaction space. Here, $G_\pm$ can be understood as the chiral transition matrix elements $\frac{G_x \mp iG_y}{\sqrt{2}}$.
It should be noted that $G_\mp$ couples with $Q_\pm$ (see Supplementary Material \cite{Supplemental}). We have that PT on phonons gives
\begin{equation}
    OQ_\pm O^{-1} = \xi Q_\mp
\end{equation}
where $\xi$ collects any resultant phase factors. Our full $PT$ symmetry condition is given by
\begin{equation}
    H'_{\Gamma K} = O\boldsymbol{\Psi}_f^\dagger(G_-Q_+ + G_+Q_-)\boldsymbol{\Psi}_iO^{-1}+\mathrm{h.c.}
\end{equation}
which becomes
\begin{equation}
    H'_{\Gamma K} = O\boldsymbol{\Psi}_f^\dagger O^{-1}(G^*_-OQ_+ + G^*_+OQ_-)O^{-1}O\boldsymbol{\Psi}_iO^{-1}+\mathrm{h.c.}
\end{equation}
where $PT$ is antilinear so we take the complex conjugates of $G$. On the electron spinors, we have that the constructed subspace is closed under $PT$. Therefore, $PT$ on our electron spinor components is represented
by a unitary matrix
\begin{equation}
    O\boldsymbol{\Psi}_\alpha O^{-1} = U_\alpha\boldsymbol{\Psi}_\alpha,
\end{equation}
for $\alpha = f,i$. We thus have that
\begin{widetext}
\begin{equation}
    \boldsymbol{\Psi}_f^\dagger(G_-Q_+ + G_+Q_-)\boldsymbol{\Psi}_i+\mathrm{h.c.} = \boldsymbol{\Psi}_f^\dagger U_f^\dagger(\xi G^*_-Q_- + \xi G^*_+Q_+)U_i\boldsymbol{\Psi}_i+\mathrm{h.c.}
\end{equation}
\end{widetext}
The coefficients on each $Q_\pm$ between our original form of $H'_{\Gamma K}$ and our final form must be equal. We have the $PT$ condition on EPC given by
\begin{equation}
    G_\pm = \xi U_f^\dagger G^*_\mp U_i.
\end{equation}
It is convenient to work with the transition strength operators derived from Fermi's golden rule,
\begin{equation}
    M_{i,\pm} = G^\dagger_\pm G_\pm \qquad M_{f,\pm} = G_\pm G^\dagger_\pm.
\end{equation}
It can be readily shown from our $PT$ condition on $G$ that $PT$ on $M$ gives
\begin{equation}
    \label{eq:PT_M}
    M_{\alpha, \pm} = U_\alpha^\dagger M^*_{\alpha, \mp} U_\alpha.
\end{equation}
This immediately shows that $M_{\alpha, +}$ and $M_{\alpha, -}$ have the same eigenvalues. Let us further say that $M_{\alpha, +}$ has eigenvectors $|1\rangle, |2\rangle$ with eigenvalues $\lambda_1, \lambda_2$.
We would like for these eigenvectors to emphasize the bright/dark nature of these channels (as will be evinced later), so we will choose $\lambda_1 \ge \lambda_2$. Let $a \in \{1,2\}$ (we here derive that the pseudospin label $a$
is also appropriate to label the transition eigenvectors), then we have that
\begin{equation}
    M_{\alpha, +} | a \rangle = \lambda_a | a \rangle.
\end{equation}
We can rewrite this as (note that $M$ is hermitian so $\lambda_a$ is real)
\begin{equation}
    U_\alpha^\dagger M^*_{\alpha, +} U_\alpha(U_\alpha^\dagger | a \rangle^*) = \lambda_a (U_\alpha^\dagger | a \rangle^*),
\end{equation}
which, applying Eq. \eqref{eq:PT_M}, leads to
\begin{equation}
    M_{\alpha, -}(U_\alpha^\dagger | a \rangle^*) = \lambda_a (U_\alpha^\dagger | a \rangle^*).
\end{equation}
Given that $U_\alpha$ is defined to transform the spinor of annihilation operators, the representation of $PT$ on our electronic states is
\begin{equation}
    O|a\rangle = U_\alpha^\dagger |a\rangle^*.
\end{equation}
That is, the corresponding eigenvectors of our two transition channels are (orthogonal) $PT$ partners. So, $|a\rangle, |b\rangle$ from the definition of the electronic pseudospinor can now be fully constrained as the eigenvectors of the transition matrices which
themselves have been shown to be $PT$ partners. Importantly, these two $PT$-partner electronic states couple to $PT$-partner phonon channels with the same transition eigenvalue (Fig.~\ref{fig:2H-transition}). Explicitly, that means that the state which $Q_+$ brightly couples with
is the $PT$ partner of the state which $Q_-$ brightly couples with, and their bright coupling strengths are symmetrically equal. It remains to show that $C_3$ symmetry indeed enforces $a,b$ selectivity with respect to phonon helicity at the ($C_3$-invariant) $K$, $K'$ points. Since the operators $Q_\pm$ have $C_3$ PAM $\pm 1$ by construction, let us say they acquire
eigenvalues $\chi_\pm$ under $C_3$. Let $R_f$, $R_i$ be the (unitary) representations of $C_3$ on the final and initial 2D degenerate electron doublets. $C_3$ invariance on the Hamiltonian, similar to
the $PT$ invariance we derived above, which can be interpreted as conservation of PAM, allows us to readily show that
\begin{equation}
    R_f^{-1} G_\pm R_i = R_f^{\dagger} G_\pm R_i = \chi_\pm G_\pm.
\end{equation}
Then, the symmetry structure easily follows as the commutator relation
\begin{equation}
    [R_\alpha, M_{\alpha,\pm}] = 0.
\end{equation}
So, the eigenstates of $M$ are also those of $C_3$ with conjugate $C_3$ eigenvalues (as related by the antiunitary $PT$). Therefore, conservation of $C_3$ eigenvalue across the entire EPC element produces chiral selection rules.
We note that, due to the spin-$\frac{1}{2}$ property of electrons, $R^3 = -I$ for a given electron state, meaning that its $C_3$ eigenvalues must be cubic roots of $-1$, or $\chi \in \{e^{-i\frac{\pi}{3}}, -1, e^{i\frac{\pi}{3}}\}$.
Thus, for a $PT$-related doublet whose $C_3$ eigenvalues are conjugates, we either have that $\chi_1, \chi_2$ be the complex members of that set, or that $\chi_1 = \chi_2 = -1$. There are three unique cases of which each produces qualitatively distinct
outcomes. First, if both the initial and final doublets are $C_3$-real (not observed in our computational results), then $C_3$ eigenvalue conservation forces chiral phonon coupling to be zero. Then, only linear phonons are allowed to couple with
the electrons since their transitions are achiral. However, if one of the doublets is $C_3$-complex, and the other remains $C_3$-real (e.g. in the CB of $2H$-bilayer MoS$_2$), then only chiral not linear phonons are allowed to couple with the electrons
since the transitions are all chiral, in which the chiral phonons will helically select $PT$-partners. Finally, if both doublets are $C_3$-complex (e.g. in the VB of $2H$-bilayer MoS$_2$), then both linear and chiral phonons are allowed to couple with
the electrons, with chiral phonons still exhibiting helical $PT$-partner selectivity. In any case, helical selectivity of $PT$-partners is observed, and we can therefore have $\lambda_2=0$ and $\lambda_1 = \lambda$ as a single chiral coupling strength (it may be $0$ in the case of two $C_3$-real doublets).

For a transition between a $C_3$-real and $C_3$-complex doublet, we note that on the given $C_3$-real doublet, $C_3$ acts as $R=-I$, meaning that for $C_3$-real doublets, the transition eigenstates are not constrained by $C_3$, since all states in that doublet will be $C_3$ eigenstates with the same $C_3$ eigenvalue of $-1$.
This means that the $PT$-partner transition eigenstates in a $C_3$-real doublet have a degree of freedom independent of $C_3$ symmetry; the selected bright transition eigenstates may even be different across phonon modes. We expect that similar rules and phenomena apply for $PT$ $C_n$ materials beyond $C_3$ by considering the interplay of $C_n$ with $PT$ as we have done for $C_3$.

\begin{figure}
    \centering
    \includegraphics[width=\linewidth]{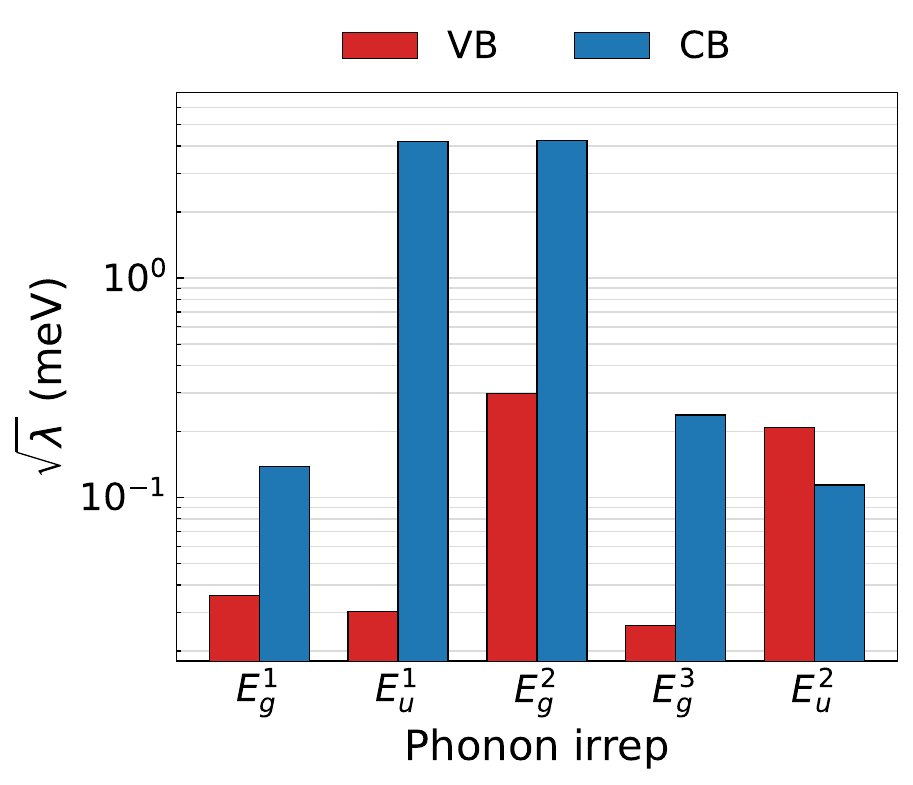}
    \caption{Chiral coupling strengths ($\sqrt{\lambda}=\sqrt{\lambda_1}$) for each phonon mode on the doubly degenerate electronic transition spaces in both the conduction (blue) and valence (red) band, plotted on a logarithmic (base-$10$) scale.
    $\sqrt{\lambda_2}$ values were all found to be below $3.1 \times 10^{-4}$ meV and thus have been considered negligible.}
    \label{fig:2H-EPC}
\end{figure}

\textit{$PT$-partner selection in $2H$-bilayer MoS$_2$} -- Fig. \ref{fig:2H-transition} gives the band structure of $2H$-bilayer MoS$_2$ and a schematic of the chiral interaction within the $PT$-based electron-phonon structure for the conduction band at the electronic $K$-point.
Chiral phonons exhibit helical selectivity with respect to $PT$-partners, whose states may differ if $C_3$ acts identically on them. Note that the square root of $\lambda$ is adopted as the magnitude of our EPC constants. Fig. \ref{fig:2H-EPC} shows the chiral coupling strength between the doubly degenerate $\Gamma$-phonon and $K$-(doublet-doublet) electron transition for the interaction spaces located at the doubly degenerate conduction band (CB) and valence band (VB), split by SOC, for $2H$-bilayer MoS$_2$.
$\sqrt{\lambda_2}$ values were found to be negligible across all interactions, showing the phonon mode's selectivity within the $PT$-induced doubly degenerate electron space. On the other hand, $\sqrt{\lambda_1} = \sqrt{\lambda}$ demonstrated significant electron-phonon
coupling, similar to values found in monolayer MoS$_2$ \cite{Mustafa_2025}.

\begin{figure}
    \centering
    \includegraphics[width=\linewidth]{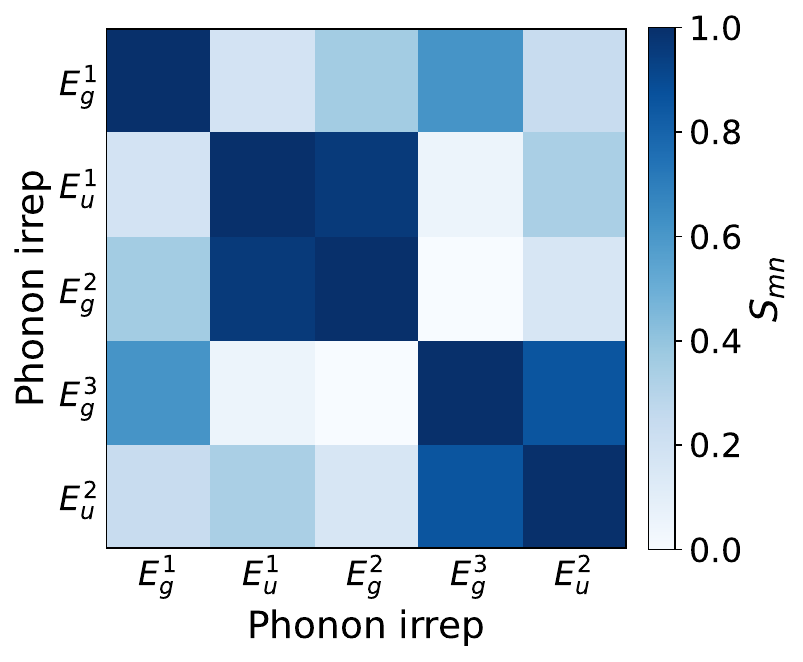}
    \caption{Overlap matrix for the transition eigenvector states between phonon modes for the initial CB electronic doublet. The values are noninteger, indicating the phonon mode selectivity of different initial transition eigenvector states in $C_3$-real spaces.}
    \label{fig:overlap}
\end{figure}

Moreover, electron doublet spaces that have real $C_3$ eigenvalues are expected to be able to demonstrate phonon mode dependence on the initial transition
eigenvectors. To show this, we have calculated values for the overlap matrix $S_{mn} = \frac{1}{2}(|\langle 1, m| 1, n\rangle|^2 + |\langle 2, m| 2, n\rangle|^2)$, shown in Fig. \ref{fig:overlap}, where $m$, $n$ label 2D phonon irreps, for $2H$-bilayer MoS$_2$'s $C_3$-real initial CB electron doublet.
Values of $1$ correspond to complete overlap and indicate that the electronic states are the same (up to an arbitrary phase factor), but we see here that overlap values (except from $m = n$) are noninteger, meaning that the phonon modes do not select the same initial electron states. Moreover,
we found that $|\langle 1 |PT| 2 \rangle|^2 = 1$ in all cases, showing the $PT$-partner nature of our selected states.

\textit{Experimental Considerations} -- Our results directly suggest that chiral phonons can be used to resolve electronic states in $PT$ materials, within the same momentum valley, via a $PT$-pseudospin mechanism. This is distinct from the well-known valley pseudospin mechanism, where phonons select different valleys' electrons, since here, the momentum degree of freedom remains constant.
Thus, we expect that in $2H$-bilayer MoS$_2$, one could drive the $Q_\pm$ phonon modes (via, e.g., circularly polarized IR on the $E^1_u$ phonon)
with optically prepared lower-CB electrons, and measure the helicity-resolved optical responses for the resultant upper-CB electrons.
The difference in the helicity-resolved responses is expected to reverse upon phonon helicity reversal, at the same momentum. Moreover, the $PT$-partner nature of the
electronic states flip spin and layer polarization upon helicity reversal in both doublets, including the $C_3$-real initial doublet in the conduction band. Furthermore, for this initial doublet in the CB, switching phonon modes produces different
spin and layer polarization expectation values due to their different wavefunctions.

\textit{Summary} -- We have studied the electron-chiral phonon interaction in the $PT$-symmetric $2H$-bilayer MoS$_2$. We show that chiral phonons separate the $PT$-induced doubly degenerate electronic space into $PT$-partner transition
channels labeled by a $PT$-induced band pseudospin. Applying $C_3$ symmetry to this interaction demonstrates helical selectivity on these transition channels in which opposite helicities select one of the $PT$-partner channels.
We expect that similar methods generalize this phenomenon to other $PT$ $C_n$ systems. This allows us to predict chiral phonons as resolving intravalley electronic states in $PT$ materials.

\textit{Acknowledgement} -- J.A.Y. acknowledges support from the School of Emerging Technology (SET) and the Fisher College of Science and Mathematics (FCSM) at the Towson University.

\bibliography{references}

\pagebreak
\appendix{
\widetext
\begin{center}
\textbf{\large Supplementary Information for\\
``Helicity-Selective Electron-Phonon Coupling Between $PT$-Partner States in $2H$-bilayer MoS$_2$"}
\end{center}
%%%%%%%%%% Merge with supplemental materials %%%%%%%%%%
%%%%%%%%%% Prefix a "S" to all equations, figures, tables and reset the counter %%%%%%%%%%
\setcounter{equation}{0}
\setcounter{figure}{0}
\setcounter{table}{0}
\setcounter{page}{1}
\makeatletter
\renewcommand{\theequation}{S\arabic{equation}}
\renewcommand{\thetable}{S\arabic{table}}
\renewcommand{\thefigure}{S\arabic{figure}}
\renewcommand{\bibnumfmt}[1]{[S#1]}
\renewcommand{\citenumfont}[1]{S#1}
%%%%%%%%%% Prefix a "S" to all equations, figures, tables and reset the counter %%%%%%%%%%
\section{Phonon physics and chirality}
\label{sec:theory}
Here we will review the relevant background of phonon physics and phononics towards the chiral phonons.
\subsection{Phonon polarization and mechanical angular momentum}
The phonon modes are derived semiclassically as small vibrational displacements of the crystal lattice. Let the $\kappa$-th atom in unit cell $i$ have position vector $\mathbf{R}_i + \tau_\kappa + \mathbf{u}_{i\kappa}$.
Here, $\mathbf{R}_i$ translates the unit cells of the crystal lattice, and $\tau_\kappa$ gives the position of the $\kappa$-th atom in the unit cell.
By definition in the relaxed crystal, $\mathbf{u}_{i\kappa} = 0$ is at an energy minimum. Admitting only the second order in energy, one derives the coupled harmonic oscillators
\begin{equation}
    m_\kappa \ddot{u}_{i\kappa\alpha} = -\sum_{j\kappa'\beta}\left.\frac{\partial^2 V}{\partial u_{i\kappa\alpha}\partial u_{j\kappa'\beta}}\right|_{\mathbf{u}=0}u_{j\kappa'\beta},
\end{equation}
where the greek letters $\alpha, \beta$ label Cartesian indices. The solution is easily obtained in mass-weighted Fourier space with the transformation
\begin{equation}
    u_{i\kappa\alpha}(t) = \frac{1}{\sqrt{N m_\kappa}}\sum_{\mathbf{q}\nu}Q_{\mathbf{q}\nu}(t)\epsilon_{\mathbf{q}\nu\kappa\alpha}e^{i\mathbf{q}\cdot(\mathbf{R}_i+\tau_\kappa)},
\end{equation}
absorbing the time dependence into the displacement coordinate. Defining the dynamical matrix as
\begin{equation}
    D_{\kappa\alpha,\kappa'\beta}(\mathbf{q}) = \frac{1}{\sqrt{m_\kappa m_{\kappa'}}}\sum_j\frac{\partial^2 V}{\partial u_{0\kappa\alpha}\partial u_{j\kappa'\beta}}|_{\mathbf{u}=0}e^{i\mathbf{q}\cdot(\mathbf{R}_j+\tau_{\kappa'}-\tau_\kappa)},
\end{equation}
we see that the dispersion relation $\omega_{\mathbf{q}\nu}$ and polarization $\boldsymbol{\epsilon}_{\mathbf{q}\nu}$ are obtained from the eigenvalue problem
\begin{equation}
    \sum_{\kappa'\beta}D_{\kappa\alpha,\kappa'\beta}(\mathbf{q})\epsilon_{\mathbf{q}\nu\kappa'\beta} = \omega_{\mathbf{q}\nu}^{2}\epsilon_{\mathbf{q}\nu\kappa\alpha}.
\end{equation}
The phonon field is then obtained by imposing quantization on the normal modes. Given the momentum operator $\Pi_{\mathbf{q}\nu} = \dot{Q}_{-\mathbf{q}\nu}$, we write down the Hamiltonian
\begin{equation}
    H = \frac{1}{2}\sum_{\mathbf{q}\nu}(\Pi_{\mathbf{q}\nu}\Pi_{-\mathbf{q}\nu} + \omega_{\mathbf{q}\nu}^2 Q_{\mathbf{q}\nu}Q_{-\mathbf{q}\nu}).
\end{equation}
Imposing the canonical commutation relations $[Q_{\mathbf{q}\nu}, \Pi_{\mathbf{q'}\nu'}] = i\hbar \delta_{\mathbf{q}\mathbf{q'}}\delta_{\nu\nu'}$, the phonon annihilation operator is
\begin{equation}
    b_{\mathbf{q}\nu} = \sqrt{\frac{\omega_{\mathbf{q}\nu}}{2\hbar}}(Q_{\mathbf{q}\nu}+i\frac{\Pi_{-\mathbf{q}\nu}}{\omega_{\mathbf{q}\nu}}).
\end{equation}
The quantum harmonic oscillator Hamiltonian is then solved as
\begin{equation}
    H = \sum_{\mathbf{q}\nu}\hbar\omega_{\mathbf{q}\nu}(b^\dagger_{\mathbf{q}\nu}b_{\mathbf{q}\nu} + \frac{1}{2}).
\end{equation}
Note that, in the main Letter, a dimensionless phonon displacement coordinate $Q = b + b^\dagger$ is used for convenience.
\subsection{Pseudo-angular momentum and selection rules}
Mechanical angular momentum is derived from the motion of the phonons themselves, as done in Ref. \cite{Zhang_2014}. A distinct angular momentum called pseudo-angular momentum (PAM) is derived from the symmetries of the crystal.
If the little group of $\mathbf{q}$ has $C_n$ symmetry, the PAM eigenvalue $\ell_{\mathbf q\nu}$, given by
\begin{equation}
  \widehat C_n\boldsymbol{\epsilon}_{\mathbf q\nu} = e^{-i\frac{2\pi}{n}\ell_{\mathbf q\nu}}\boldsymbol{\epsilon}_{\mathbf q\nu},
\end{equation}
is conserved modulo $n$. At generic crystal momenta, or when the crystal does not contain such rotational symmetry, PAM is not necessarily well-defined, nor conserved.
This gives us the chiral selection rules, exhibited in electron-phonon coupling. One can readily show the following form of the chiral EPC Hamiltonian, where $X$ is an electron transition operator,
\begin{equation}
    H_\mathrm{EPC} = g_+ X Q_- + g_- X Q_+ = \frac{g_x - ig_y}{\sqrt{2}}X\frac{Q_x + iQ_y}{\sqrt{2}} +  \frac{g_x + ig_y}{\sqrt{2}}X\frac{Q_x - iQ_y}{\sqrt{2}}= g_xXQ_x + g_yXQ_y,
\end{equation}
Chiral EPC means that $g_\mp \neq 0$ only if the change in electron angular momentum is equal to the absorbed phonon PAM or opposite to the emitted phonon PAM.
\section{Full DFT results for $2H$-bilayer MoS$_2$'s $\Gamma$-point phonons}
Table \ref{tab:2H_Gamma} shows our complete DFT calculation for the $\Gamma$-point phonons in $2H$-bilayer MoS$_2$. The modes are labeled by their energy orderings, where doubly degenerate modes are grouped together.
We see that indeed the electronic space organizes itself into bright and dark coupling channels
for the chiral phonons only, whereas for the linear phonons, the two coupling strengths were found to be essentially equal and thus reported as a single coupling parameter. In $2H$, the valence band transition is between two $C_3$-complex doublets, which
allows for both helicity-selective chiral phonon interactions (selecting the transition that conserves total $C_3$ eigenvalue and conjugates electronic $C_3$ eigenvalue), and unselective lienar mode coupling (that leaves electronic $C_3$ eigenvalue invariant).
The conduction band transition, however, is between $C_3$-real and $C_3$-complex spaces, which must be chiral. Hence, linear phonons cannot interact with these electrons, but chiral phonons can.
\begin{table}
    \centering
    \caption{$2H$-bilayer MoS$_2$ phonons at $\Gamma$. We have separated them between the linear and degenerate phonons. For the degenerate phonons,
    we have considered the bright/dark channels due to the chiral combinations in EPC. For the linear phonons, we note that the two transition strength eigenvalues square roots were found to be equal up to
    $2.3 \times 10^{-4}$ meV, so we have reported their shared value $\sqrt{\lambda} = \sqrt{\lambda_1} = \sqrt{\lambda_2}$. Only the valence band exhibits coupling to the linear modes, explained by $C_3$ eigenvalues. All $0$ values reported were below $3.1 \times 10^{-4}$ meV.}
    \label{tab:2H_Gamma}
    \begin{tabular}{ccccc}
        \hline\hline
        Modes & Energy (meV) & $\sqrt{\lambda^\mathrm{VB}_{1(2)}}$ (meV) & $\sqrt{\lambda^\mathrm{CB}_{1(2)}}$ (meV) \\
        \hline
        $4/5$   & $3.0541$ & $0.03591 (0)$ & $0.13866 (0)$ \\
        $7/8$   & $34.747$ & $0.03024 (0)$ & $4.1780 (0)$ \\
        $9/10$  & $34.937$ & $0.29849 (0)$ & $4.2526 (0)$ \\
        $11/12$ & $46.967$ & $0.02617 (0)$ & $0.2383 (0)$ \\
        $13/14$ & $46.981$ & $0.20832 (0)$ & $0.11399 (0)$ \\
        \hline\hline
        Modes & Energy (meV) & $\sqrt{\lambda^\mathrm{VB}}$ (meV) & $\sqrt{\lambda^\mathrm{CB}}$ (meV) \\
        \hline
        $6$     & $5.3269$ & $4.404$ & $0$ \\
        $15$    & $49.645$ & $22.068$ & $0$ \\
        $16$    & $50.173$ & $2.8719$ & $0$ \\
        $17$    & $57.601$ & $12.314$ & $0$ \\
        $18$    & $57.830$ & $5.5283$ & $0$ \\
        \hline\hline
    \end{tabular}
\end{table}

\section{Spin and layer polarizations of transition eigenstates}
\begin{table}
    \centering
    \renewcommand{\arraystretch}{1.15}
    \caption{Spin ($S_z$) and layer ($P_L$) polarization values for the initial and final bright transition eigenstates in $2H$-bilayer MoS$_2$ across phonon modes for combination $Q_\pm$. These values were
    calculated based on the Wannierization provided by QE. Indeed, their nonzero polarizations reverse upon phonon helicity switching, and such values are different for the initial CB states since there is no
    $C_3$ constraint on the transition eigenstates.}
    \label{tab:spin_layer}
    \begin{tabular}{c|ccccc}
        \hline\hline
        VB & $4/5$ & $7/8$ & $9/10$ & $11/12$ & $13/14$ \\
        \hline
        $S_{z,i}/\hbar$ & $\mp0.4992$ & $\mp0.4992$ & $\mp0.4992$ & $\mp0.4992$ & $\mp0.4992$ \\
        $P_{L,i}$       & $\pm0.8260$ & $\pm0.8260$ & $\pm0.8260$ & $\pm0.8260$ & $\pm0.8260$ \\
        $S_{z,f}/\hbar$ & $\pm0.4999$ & $\pm0.4999$ & $\pm0.4999$ & $\pm0.4999$ & $\pm0.4999$ \\
        $P_{L,f}$       & $\pm0.8256$ & $\pm0.8256$ & $\pm0.8256$ & $\pm0.8256$ & $\pm0.8256$ \\
        \hline
        CB & $4/5$ & $7/8$ & $9/10$ & $11/12$ & $13/14$ \\
        \hline
        $S_{z,i}/\hbar$ & $\pm0.3172$ & $\mp0.4987$ & $\mp0.4594$ & $\pm0.4465$ & $\pm0.1662$ \\
        $P_{L,i}$       & $\pm0.6349$ & $\mp0.9992$ & $\mp0.9206$ & $\pm0.8947$ & $\pm0.3336$ \\
        $S_{z,f}/\hbar$ & $\pm0.4938$ & $\pm0.4938$ & $\pm0.4938$ & $\pm0.4938$ & $\pm0.4938$ \\
        $P_{L,f}$       & $\mp0.9985$ & $\mp0.9985$ & $\mp0.9985$ & $\mp0.9985$ & $\mp0.9985$ \\
        \hline\hline
    \end{tabular}
\end{table}
In table \ref{tab:spin_layer}, we have given our values for the spin and layer polarizations of the transition eigenstates for both the initial and final electronic spaces of both the conduction and valence bands
with respect to phonon mode doublets with combination $Q_\pm$.
We calculated these values from the Wannierization provided by QE in which the layer polarization was defined to be
\begin{equation}
    \hat{P}_L = \sum^n_i \epsilon_i |i\rangle\langle i |
\end{equation}
for our $n$ Wannier functions (in our case, $n = 44$), where each $\epsilon_i$ for Wannier function $|i\rangle$ was defined to be $\pm 1$ depending on if its centre was above or below the middle of our bilayer. We see
that each state is both spin and layer polarized whose polarization flips upon phonon helicity reversal. Notably, in the conduction band, we see that the polarizations of the bright initial transition state varies across phonon modes
reflecting that these states live in a $C_3$-real space.
%\bibliography{references}
\end{document}